\documentclass[nofootinbib,showpacs,preprintnumbers,amsmath,amssymb,floatfix]
{revtex4-1}
\usepackage{epsf}

\begin{document}


\title{Origin of the $Q^2$-dependence of the DIS structure functions}

\vspace*{0.3 cm}

\author{B.I.~Ermolaev}
\affiliation{Ioffe Physico-Technical Institute, 194021
 St.Petersburg, Russia}
\author{M.~Greco}
\affiliation{Department of Physics and INFN, University Rome III,
Rome, Italy}
\author{S.I.~Troyan}
\affiliation{St.Petersburg Institute of Nuclear Physics, 188300
Gatchina, Russia}

\begin{abstract}
We consider in detail the $Q^2$ -dependence of the DIS structure functions, with $Q$ being the virtual photon momentum. Very often this
dependence is claimed to be originated by the $Q^2$-dependence of the QCD coupling. This leads to
the small-$x$ asymptotics of the structure functions with $Q^2$ -dependent intercepts. We demonstrate that the
DGLAP parametrization $\alpha_s = \alpha_s (Q^2)$ is an approximation valid in the region of large $x$ (where $2pq$ can
be approximated by $Q^2$)
only, providing the
factorization scale is also large.  Outside this region, the DGLAP parametrization fails,
so $\alpha_s$ should be replaced by an effective coupling which is independent of $Q^2$ at small $x$.
As a consequence, intercepts of the structure functions  are independent
of $Q^2$. Nevertheless, the small-$x$ asymptotics of the structure functions
explicitly depend on $Q^2$, even when the coupling does not depend on it.
We also consider the structure functions at small $Q^2$ and give a
comment on power-$Q^2$ corrections to the structure functions at large and small $Q^2$.
\end{abstract}


\maketitle

\section{Introduction}

Each of the DIS structure functions has two independent arguments: the invariant energy $w = 2pq$
(with $p$ and $q$ being the hadron target and virtual photon momenta respectively) and
virtuality of the incoming photon $Q^2 = - q^2 > 0$. Early experiments of 60s at SLAC proved that the
structure functions depended on only
one argument $x = Q^2/w$. This phenomenon was called scaling. Later, new experiments proved that the scaling
was violated i.e. in addition to $x$-dependence the
DIS structure functions also depended on $Q^2$ straightforwardly. Since that it become conventional to regard variables
$x$ and $Q^2$ as independent arguments of the structure functions. However, one has to realize that
those arguments (in contrast to the set of really independent arguments $w, Q^2$) cannot be regarded
as independent variables at all times.
Indeed, they are independent when $Q^2$ is fixed
while $w$ is scanned but they are not independent at fixed $w$ and scanned $Q^2$. The $Q^2$ -dependence of the structure functions
is the objective of the DGLAP evolution equations\cite{dglap} implementing the total resummation of
$\ln^k Q^2$.  In DGLAP and beyond this approach,  the $Q^2$ -dependence of the structure functions is often
thought to be induced by the $Q^2$ -dependence of $\alpha_s$:

\begin{equation}\label{adglap}
\alpha_s =  \alpha_s (Q^2) .
\end{equation}

As this parametrization takes place in the DGLAP equations\cite{dglap}, throughout the present paper we will
address it as the DGLAP-parametrization.
Parametrization (\ref{adglap}), originally deduced for large $x$, where $2pq$ can be approximated by $Q^2$,
is often exploited to describe DIS in the
small-$x$ region and even in the small-$x$ asymptotic of the structure functions, i.e. at
$x \to 0$ (see e.g. Refs.~\cite{bad}-\cite{ross}). As a result, the small-$x$ asymptotics look like

\begin{equation}\label{asq}
f \sim x^{- \Delta (Q^2)},
\end{equation}
with the $Q^2$-dependence of the asymptotics originated by the "intercept" $\Delta (Q^2)$.

In the present paper we argue against both validity of Eq.~(\ref{asq}) and treating Eq.~(\ref{adglap})
as an exact formula which could be used at any $x$ and $Q^2$.
In what follows we will show that the parametrization (\ref{adglap}) should be used with a certain care:
it can be used within the proper DGLAP kinematic region, i.e. at large $x$ and $Q^2$, however
it is not enough: the factorization scale should also be large.
We show that if any of these requirements is violated, $\alpha_s (Q^2)$
should be replaced by the effective coupling which does not depend on $Q^2$ at small $x$ as well as at
small $Q^2$. The effective coupling, being independent of $Q^2$ at small $x$,
cannot bring any $Q^2$-dependence to
the structure functions. In this case the $Q^2$ -dependence of the
structure functions is originated by alternative sources.
We also argue against regarding the approximative parametrization (\ref{adglap}) as
the basis for the shift
$\alpha_s (Q^2) \to \alpha_s (Q^2 + \mu^2)$ as a continuation into the small-$Q^2$ region
when the shift is used in order to keep $\alpha_s$
in the perturbative domain.  Instead we suggest an alternative approach to describe the
structure functions in the small-$Q^2$ region.

Our paper is organized as follows: In Sect.~II we consider the DIS structure functions
in general, focusing on their conventional
dependence on $Q^2$. In Sect.~III we consider the structure functions in the framework of
DGLAP and discuss parameterizations of the QCD coupling in different kinematic regions.
In Sect.~IV we derive a general expressions for the structure functions at small $x$ and
derive a general form of their small-$x$ asymptotics, demonstrating that there is no
room for Eq.~(\ref{asq}) at $x \to 0$. In Sect.~V we discuss the role of singular
factors conventionally used in DGLAP-fits for initial parton distributions. We
consider the structure functions at small $Q^2$ in Sect.~VI. We make a brief 
remark on higher twists in Sect.~VII. 
 Finally, Sect.~VIII is for our concluding remarks.

\section{General structure of the DIS structure functions}

All DIS structure functions are given by a convolution of perturbative components
$f^{(pert)}_r$ and parton distributions $\Phi_r$. For instance, in collinear factorization and
when DGLAP is used, the non-singlet structure function $f_{NS} (x, Q^2)$ is represented as the convolution

\begin{equation}\label{colconvns}
f_{NS} (x, Q^2) =  C_q (x,x_0) \otimes \Delta q (x_0, Q^2)
\end{equation}
where $C_q$
is the quark
coefficient function while
$\Delta q (Q^2)$  stands for the quark evolved distribution defined at scale $Q^2$.
$C_q (x, x_0)$ controls evolution of this distribution from $x_0$ to $x$.
The $Q^2$ -evolution of the quark distribution from the initial scale $\mu^2$ to scale $Q^2$ is controlled by the DGLAP
evolution equations\cite{dglap}. It can be written as

\begin{equation}\label{dglapgen}
\Delta q (x_0, Q^2) = E(Q^2, \mu^2) \otimes \delta q (x_0, \mu^2)
\end{equation}
where the initial quark distribution $\delta q (x_0, \mu^2)$  is
conventionally chosen at $x_0 \sim 1$ and $\mu \sim 1$ GeV.   The scale $\mu^2$
is called the factorization scale.
Eqs.~(\ref{colconvns}, \ref{dglapgen}) can be written as one convolution:

\begin{equation}\label{fpert}
f_{NS} (x, Q^2) =  \left[C_q (x,y) \otimes E(Q^2, \mu^2) \right] \otimes \delta q (\mu^2) \equiv
f_{NS}^{(pert)} (x, Q^2)\otimes \delta q (\mu^2)
\end{equation}
where $f_{NS}^{(pert)}$ denotes the whole perturbative evolution of $\delta q$ both with respect to $x$ and $Q^2$.
Eq.~(\ref{fpert}) can easily be generalized to the case of the singlet
structure functions by adding gluon contributions. So, both in basic factorization\cite{egtfact} as well as in
collinear and
$k_T$ -factorizations, such a convolution can be written as

\begin{equation}\label{convgen}
f (x, Q^2) = \sum_r f^{(pert)}_r \left(q\kappa, Q^2, \kappa^2\right) \otimes \Phi_r \left(p\kappa, \kappa^2\right),
\end{equation}
where $\Phi$ denotes parton distributions in any of the factorizations, the subscript $r$ marks the intermediate partons (quarks and qluons) with momentum $\kappa$
and the summation over $r$ takes place when necessary. The notation $\otimes$ means the one-dimensional integration
over the longitudinal for collinear factorization and two-dimentional (both the longitudinal and
transverse momenta) in the case of $k_T$ -factorization; $\kappa^2 = \mu^2$ in the case of collinear factorization.
In Eq.~(\ref{convgen}) we have dropped a possible
dependence on unessential variables like the target spin and mass.

\subsection{General structure of the perturbative component}

The simplest form of structure functions corresponds to collinear factorization where
it can be written in terms the Mellin integral:

\begin{equation}\label{fgen}
f = \int^{\imath \infty}_{- \imath \infty} \frac{d \omega}{2 \pi \imath}
x^{- \omega}
\left[ f^{(pert)}_q (\omega) \delta q (\omega)  + f^{(pert)}_g (\omega)
\delta g (\omega),
\right]
\end{equation}
with $\delta g$ and $\delta g (\omega)$ standing for the quark and gluon distributions in the $\omega$ -space
respectively.
In Eq.~(\ref{fgen}) they are convoluted with
the perturbarive contributions $f^{(pert)}_r$. These perturbative contributions in any factorization have
the following generic structure:

\begin{equation}\label{fpertgen}
f^{(pert)}_r = \int^{\imath \infty}_{- \imath \infty} \frac{d \omega}{2 \pi \imath}
x^{- \omega} C_{r} (\omega, \alpha_s, \kappa^2) e^{\widetilde{\Omega}_{r} (\omega, \alpha_s, Q^2, \kappa^2)}
\end{equation}
where the term $C_r$ is a generic notation for the coefficient functions and $\widetilde{\Omega}_r$ is
expressed through the matrix of anomalous dimensions.
For instance  in LO DGLAP, where, in collinear factorization, $\kappa^2 = \mu^2$, the non-singlet perturbative
component $f^{(pert)}_{NS} (\equiv f^{(pert)}_q)$ is expressed
in terms of the non-singlet coefficient function $C_{NS}$ and $\widetilde{\Omega}_{NS}$:

\begin{equation}\label{cgammalo}
C^{(LO)}_{NS} = 1,~~~ \Omega^{(LO)}_{NS} = \gamma^{(0)}_{qq} (\omega) \int_{\mu^2}^{Q^2}
\frac{d k^2_{\perp}}{k^2_{\perp}} \alpha_s (k^2_{\perp})  = \gamma^{(0)}_{qq} (\omega) \frac{1}{b}
\ln \left[\frac{\ln \left(Q^2/\Lambda^2\right)}{\ln \left(\mu^2/\Lambda^2\right)}\right],
\end{equation}
with $\gamma^{(0)}_{qq} (\omega)$ being the well-known quark-quark anomalous dimension and $b = (33 - 2 n_f)/(12 \pi)$.
In NLO DGLAP, both $C_{NS}$ and $\Omega_{NS}$ acquire additional terms proportional to $\alpha_s$.
In particular, the structure of the NLO coefficient functions is

\begin{equation}\label{cnlo}
C^{(NLO)}_{NS} = 1 + \alpha_s (Q^2) \tilde{C}_{NS} (\omega).
\end{equation}

We have used this simple example in order to demonstrate explicitly that
the $Q^2$ -dependence of any structure function $f$ is achieved through the $Q^2$ -dependence of
its perturbative components $f^{(pert)}_r$.
The $Q^2$ -dependence of $f^{(pert)}_r$ comes from the upper limit $Q^2$ of integrations
over $k_{\perp}^2$ in expressions for both $\Omega_r$, as explicitly shown in Eq.~(\ref{cgammalo}),
and in expressions for coefficient functions. The latter is related to the parametrization
$\alpha_s =  \alpha_s (Q^2)$.
This feature is common for all structure functions, although
specific expressions for $C_{r r'}$ and $\Omega{r r'}$ are different for different structure functions.

\section{$Q^2$ -dependence of the QCD coupling}

In this Sect. we discuss parameterizations of $\alpha_s$ in DGLAP equations in collinear factorization
at different values of $x$ and $\mu^2$
Although for the sake of simplicity we consider the LO DGLAP equations, our conclusions are valid for NLO, NNLO and so on.
In the LO DGLAP evolution equation the parametrization (\ref{adglap})
appears when the integral DGLAP equations for $f^{(pert)}$

\begin{equation}\label{dglapint}
f^{(pert)}_r (x, Q^2) =  \int_{\mu^2}^{Q^2} \frac{d k^2_{\perp}}{k^2_{\perp}}
\int_x^1 \frac{d \beta}{\beta} \alpha (k^2_{\perp})
 f^{(pert)}_{r'} (x/\beta, Q^2/k^2_{\perp}) P_{r'r} (\beta)
\end{equation}
are reduced to the differential equations

\begin{equation}\label{dglap}
\frac{\partial f^{(pert)}_r (x, Q^2)}{\partial \ln Q^2} = \alpha (Q^2)
\int_x^1 \frac{d \beta}{\beta} f^{(pert)}_{r'} (x/\beta, Q^2) P_{r'r} (\beta).
\end{equation}

Eqs.~(\ref{dglapint},\ref{dglap}) demonstrates explicitly that Eq.~(\ref{adglap}) holds when

\textbf{(i)} the upper limit of integration over $k_{\perp}^2$ in Eqs.~(\ref{dglapint})
is $Q^2$ which corresponds to $x \sim 1$.

\textbf{(ii)} the parametrization

\begin{equation}\label{ast}
\alpha_s = \alpha_s (k^2_{\perp})
\end{equation}

is used in the involved Feynman graphs. Throughout the present paper we will refer to the
parametrization of Eq.~(\ref{ast}) as the standard parametrization.
According to the results of Ref.~\cite{etalpha},  one can use the standard parametrization
(\ref{ast})  in expressions for
$f^{(pert)}$ in collinear factorization
only when the following two conditions are fulfilled:
First, values of $x$ should be large:
\begin{equation}\label{xlarge}
x \sim 1
\end{equation}
to ensure the use of $Q^2$ instead of $2pq$.
Second, the factorization point $\mu^2$ should obey the strong inequality
\begin{equation}\label{mupi}
\mu^2 \gg e^{\pi} \Lambda^2 \approx 23 \Lambda^2.
\end{equation}

When one of both of the requirements in Eqs.~(\ref{xlarge}, \ref{mupi}) are violated, the coupling $\alpha_s (k^2_{\perp})$ in
the integral
DGLAP evolutions equations for $f^{(pert)}$,
should be replaced by the effective coupling $\alpha_{eff}$, as shown in Ref.~\cite{etalpha}. This coupling can be represented by
different approximative expressions, depending on the kinematics. In Subsections~\textbf{A} and \textbf{B} we consider
the cases of violating either (\ref{xlarge}) or (\ref{mupi})
 and consider violation of both of them in Subsection~\textbf{C}.

\subsection{Case A}

When Eq.~(\ref{mupi}) is violated but $x$ is so close to $1$ that essential $\beta$ in Eq.~(\ref{dglapint}) are
$\sim 1$, the coupling $\alpha_s (k_{\perp}^2)$ in Eq.~(\ref{dglapint}) should be replaced by

\begin{equation}\label{aeffsmmu}
\alpha_{eff}(k_{\perp}^2) = \frac{1}{b} \frac{l_0}{(l^2_0 + \pi^2)} - \frac{1}{\pi b} \arctan \left(\frac{\pi}{l_0}\right)
+  \frac{1}{\pi b} \arctan \left(\frac{\pi}{\widetilde{l}(k^2_{\perp})}\right),
\end{equation}
where we have denoted
\begin{equation}\label{l}
l_0 = \ln (\mu^2/\Lambda^2),~~~ \widetilde{l}(k^2_{\perp}) = \ln (k^2_{\perp}/\Lambda^2).
\end{equation}

It leads to replacement of the integration of $\alpha_s (k^2_{\perp})$ in Eq.~(\ref{cgammalo}) by the
integration of $\alpha_{eff}((k^2_{\perp})$ over the same interval:

\begin{equation}\label{aeffbigx}
\int_{\mu^2}^{Q^2} \frac{d k_{\perp}^2}{k_{\perp}^2}~ \alpha_s (k_{\perp}^2)) \to  \int_{\mu^2}^{Q^2} \frac{d k_{\perp}^2}{k_{\perp}^2}~ \alpha_{eff} (k_{\perp}^2)).
\end{equation}

Let us notice that when $\mu$ obeys Eq.~(\ref{mupi}), the first and second terms in Eq.~(\ref{aeffsmmu}) cancel
each other while the third term can be approximated by $\alpha_s (k^2_{\perp})$. This reduces the integral in Eq.~(\ref{aeffbigx})
to the  DGLAP expression in Eq.~(\ref{cgammalo}).

\subsection{Case B}

Let us consider the opposite situation when $\mu$ satisfies Eq.~(\ref{mupi}) but $x$ is small.
In this case $\alpha_s (k_{\perp}^2)$ in Eq.~(\ref{dglapint}) should be
replaced by $\alpha_s (k_{\perp}^2/\beta)$ and the upper limit of integration over $k_{\perp}^2$ should
be changed. The point is that integration over $k_{\perp}^2$ always runs from some starting point $\mu^2$
(which is fixed from physical considerations and sometimes coincides with the factorization
scale) to
the total invariant energy
\begin{equation}\label{swx}
s = (p+q)^2 \approx w (1-x),
\end{equation}
with $w = 2pq$.
In the DGLAP framework, where $x$ is not far from $1$, $s \approx Q^2$, so in the DGLAP
equation (\ref{dglapint}) the upper limit is  $Q^2$. On the other hand, $s \approx w$ at small $x$.
This leads to replacement of Eq.~(\ref{dglapint})
by the following equation:

\begin{equation}\label{dglapintsmx}
f^{(pert)}_r (x, Q^2) =  \int_{\mu^2}^{w} \frac{d k^2_{\perp}}{k^2_{\perp}}
 \int_{\beta_0}^1 \frac{d \beta}{\beta} \alpha_s (k^2_{\perp}/\beta)
 f^{(pert)}_{r'} (x/\beta, Q^2/k^2_{\perp}) P_{r'r} (\beta).
\end{equation}
with
\begin{equation}\label{beta0}
\beta_0 = x + k^2_{\perp}/w .
\end{equation}

Obviously, $\beta_0 \approx x$ at large $x$. In addition, one can neglect the $\beta$ -dependence of
$\alpha_s (k^2_{\perp}/\beta)$ at large $x$, arriving back to the integral DGLAP equation (\ref{dglapint}).

\subsection{Case C}

Finally, when $\mu$ does not satisfy Eq.~(\ref{mupi}) and, in addition, $x \ll 1$,
the coupling $\alpha_s (k_{\perp}^2/\beta)$ in Eq.~(\ref{dglapintsmx})
should be replaced by $\alpha_{eff} (k_{\perp}^2/\beta)$:

\begin{equation}\label{aeffsmallx}
\alpha_{eff} (k_{\perp}^2/\beta) =
\frac{1}{b} \frac{l_0}{(l^2_0 + \pi^2)} - \frac{1}{\pi b} \arctan \left(\frac{\pi}{l_0}\right)
+  \frac{1}{\pi b} \arctan \left(\frac{\pi}{l}\right),
\end{equation}
with $l =  \ln \left(k_{\perp}^2/(\beta\Lambda^2)\right)$.
It converts Eq.~(\ref{dglapintsmx}) into a new equation:

\begin{equation}\label{dglapintsmxbigmu}
f^{(pert)}_r (x, Q^2) =  \int_{\mu^2}^{w} \frac{d k^2_{\perp}}{k^2_{\perp}}
\int_{\beta_0}^1 \frac{d \beta}{\beta} \alpha_s (k^2_{\perp}/\beta)
 f^{(pert)}_{r'} (x/\beta, Q^2/k^2_{\perp}) P_{r'r} (\beta).
\end{equation}

Eqs.~(\ref{dglapintsmx},\ref{dglapintsmxbigmu}) explicitly demonstrate that there is no factorization between
the integrations over $\beta$ and $k^2_{\perp}$ at small $x$. In other words, the factorization
between the longitudinal and transverse spaces taking place in DGLAP vanish in the
small-$x$ kinematics.
Eqs.~(\ref{dglapintsmx},\ref{dglapintsmxbigmu}) also show that structure functions at small $x$
 depend on $Q^2$ through the integration limit $\beta_0$ while
the effective couplings at $x \ll 1$ differ a lot from the standard parametrization given
by Eq.~(\ref{adglap}). Eq.~(\ref{dglapintsmxbigmu}) corresponds to the case when both Eq.~(\ref{xlarge}) and (\ref{mupi})
are violated while the expressions for the couplings and structure functions in Subsects. \textbf{A} and \textbf{B}
correspond to violation of either Eq.~(\ref{xlarge}) or (\ref{mupi}) and can easily be obtained from
Eqs.~(\ref{aeffsmallx},\ref{dglapintsmxbigmu}). Nevertheless, we accentuate that Eq.~(\ref{dglapintsmxbigmu}) is also
approximation obtained in Ref.~\cite{etalpha} in order to factorize the coupling. This expression
should be used only in the context of evolution equations. A more general treatment of
the QCD coupling, where the factorization was not required, was done in Ref.~\cite{egtalpha}. We discuss it briefly in the next Sect.
To conclude, we would like to notice that Eq.~(\ref{dglapintsmxbigmu}) was obtained with constructing Dispersion Relations
for forward scattering
amplitudes. This approach is similar to the approach of Ref.~\cite{ds} but
differs from the alternative approaches (see e.g. Refs.~\cite{agl,ricc}) where the Dispersion Relations
were constructed for $\alpha_s$ by itself.

Obviously,
 logarithms of $x$ are large in the small-$x$ region, so their resummation to all orders in $\alpha_s$ is important.
 This is beyond of the reach of DGLAP, where only logarithms of $Q^2$ are resummed.
 A generalization of the DGLAP equations to the small-$x$ region was done
 in Refs.~\cite{egtsns} (see also overview \cite{g1sum}) for the spin structure
 function $g_1$ and non-singlet component of $F_1$: there were resummed
the leading, double logarithms of both $x$ and $Q^2$ and at the same time
the QCD coupling in each rung of the involved Feynman graphs was running.
Applying
the saddle-point method to the structure functions obtained in \cite{egtsns} exhibited their Regge
behavior at $x \to 0$, with the intercepts being just numbers without any parameters (e.g $\alpha_s$, etc).
The intercept of the non-singlet
$F_1$ proved to be $0.38$, the intercept of the non-singlet $g_1$ was $0.42$ and the intercept
of the singlet $g_1$ was $0.86$.

\section{General structure of the evolution equations for DIS structure functions}

In this Sect. we, consider general expressions for the structure functions, skipping unessential details. Throughout
the present Sect. we imply that collinear factorization is used, though a generalization to  $k_T$
-factorization is easy to do.
As equations for the structure functions involve convolutions  (see e.g. Eq.~(\ref{fpert})),
it is convenient to represent the perturbative contribution $f_r^{(pert)}$ in terms of the Mellin
transform:
\begin{equation}\label{mellin}
f_r^{(pert)} = \int^{\imath \infty}_{- \imath \infty} \frac{d \omega}{2 \pi \imath} x^{- \omega}
F_r (\omega,y),
\end{equation}
were we have denoted $y = \ln (Q^2/\mu^2)$ and introduced the Mellin amplitude $F_r$.
Focusing on the $t$ -channel intermediate state with two partons and neglecting other
contributions  (see for detail Ref.~\cite{g1sum}) allows us to compose the
following system of equations for $F_r$:

\begin{equation}\label{eqf}
\left[\omega + \partial/ \partial y \right] F_r (\omega, y) = F_{r'}(\omega, y) H_{r'r} (\omega)
\end{equation}
 Solving these equations, we arrive at

\begin{equation}\label{solf}
f (x,Q^2) = \sum_r
\int^{\imath \infty}_{- \imath \infty} \frac{d \omega}{2 \pi \imath} x^{- \omega}
F_r (\omega,y) \delta r (\omega) = \sum_r
\int^{\imath \infty}_{- \imath \infty} \frac{d \omega}{2 \pi \imath} x^{- \omega}
C_r (\omega) e^{y \Omega_r (\omega)} \delta r (\omega)~,
\end{equation}
where $r = q,g$, so that $\delta r = \delta q, \delta g$, and
$\Omega_r$ is made of new matrix of anomalous dimensions $H_{r'r}$. Both $C_r$ and $H_{r'r}$
account for both the total resummation of logarithms of $x$ and running $\alpha_s$ effects.
As $\alpha_s$ depends on the longitudinal Sudakov variables, it participate in the Mellin transform
(see Ref.~\cite{egtalpha} for detail).
As a result, $\alpha_s$ with the time-like argument is replaced by

\begin{equation}
\label{a} A(\omega) = \frac{1}{b} \Big[\frac{\eta}{\eta^2 + \pi^2}
- \int_0^{\infty} \frac{d \rho e^{-\omega \rho}}{(\rho + \eta)^2 +
\pi^2} \Big],
\end{equation}
where $\eta = \ln(\mu^2/\Lambda^2_{QCD})$ and $\mu$ is the IR cut-off. When the argument of the coupling
is space-like, the $\pi^2$ -terms are absent.

Comparison of Eqs.~(\ref{solf}) and (\ref{fpertgen}) shows that the coefficient functions $C_r$
do not depend on $Q^2$ at all. The $Q^2$ -dependence of $f_r (x,Q^2)$ is located in the exponent of
(\ref{solf}) and it is controlled by the anomalous dimensions. The small-$x$ asymptotics of $f_r$
can be obtained with applying the saddle-point method to Eq.~(\ref{solf}). The stationary phase is
obtained as a solution to the following equation  :

\begin{equation}\label{eqomega}
\frac{d}{d \omega} \left[\omega \xi + \ln C_r (\omega) + y \Omega_r (\omega)\right] =
\xi + \frac{1}{C_r} \frac{d C_r}{d \omega} + y \frac{d \Omega_r}{d \omega} = 0,
\end{equation}
where $\xi = \ln (1/x)$.
This equation should be solved at $\xi \to \infty$ and fixed $y$ and the solution, $\omega_0$, is
called the stationary point. The rightmost root of Eq.~(\ref{eqomega}) stipulates the
small-$x$ asymptotic behavior of $f_r$, with
\begin{equation}\label{intercept}
\Delta \equiv \max [\omega_0]
\end{equation}
being the intercept of the structure function.
 Eq.~(\ref{eqomega}) implies that
the large factor $\xi$ must be equated by another large factor. As $y$ is fixed, such a factor can be
any of the following options :\\

\textbf{(i)}
\begin{equation}\label{czero}
C_r (\omega) \to - 0,
\end{equation}

\textbf{(ii)}
\begin{equation}\label{csing}
d C_r (\omega)/ d \omega \to - \infty,
\end{equation}

\textbf{(iii)}
\begin{equation}\label{hsing}
d \Omega_r (\omega)/ d \omega \to - \infty ,
\end{equation}
when $\omega \to \omega_0$.
Accounting for logarithmic contributions $\sim \ln^n x$ to $C_r (\omega)$ and $\Omega_r$ means
that they acquire contributions $\sim c_n/\omega^{1+n}$ and $ \sim c'_n/\omega^n$ respectively
and therefore any of them and $d C_r/d \omega$, which is $\sim c_n/\omega^{2 + n}$, are singular at $\omega \to \omega_0 = 0$.
 Obviously, fulfilment of Eq.~(\ref{czero}) can be achieved only when a part of $c_k$ is negative, which contradicts to calculations
made in DGLAP and to expressions for the non-singlet structure functions obtained in Ref.~\cite{egtsns}.
Then, let us note that as
contributions to $d C_r/d \omega$ are more singular than contributions to $C_r$ or $\Omega_r$,
the case \textbf{(ii)} is the most important compared to \textbf{(i}) and \textbf{(iii)}. Indeed, $\omega$ should be small in order to prevent oscillations of the factor
$\exp (\xi \omega)$ in Eq.~(\ref{solf}) at large $\xi$.
The series of the pole contributions $\sim 1/\omega^k$ can be summed up. For example, the coefficient
function $C_{NS}$ and anomalous dimension $H_{NS}$ for the non-singlet structure function $F_1^{NS}$
(for this case $\Omega_r (\omega)$ is replaced by $H_{NS} (\omega)$) proved to be

\begin{equation}\label{chns}
C_{NS}= \frac{\omega}{\omega - H_{NS}},~~~~~~H_{NS} = (1/2) \left[\omega - \sqrt{\omega^2 - B_{NS}(\omega)}\right].
\end{equation}
When $\alpha_s$ is fixed, $B_{NS} = 2 \alpha_s C_F/\pi $, otherwise it is given by a much more
complicated expression (see Ref.~\cite{egtsns} for detail). It is easy to see that all coefficients in the
expansion of $C_{NS}$ into series in $1/\omega^n$ are positive. This also excludes Eq.~(\ref{czero})
and demonstrates that Eq.~(\ref{csing}) corresponds to the most important case. So,
we arrive at the asymptotics of $f$ in the form of following contributions of the Regge type: at $x \to 0$

\begin{equation}\label{as}
f (x,Q^2) \sim  x^{-\Delta} \left(Q^2/\mu^2\right)^{\Omega (\Delta)} \delta r .
\end{equation}

Eqs.~(\ref{solf},\ref{as}) demonstrate explicitly that the dependence of the
structure functions on $Q^2$ has nothing to do with the  $Q^2$ -dependence of the coupling.

\section{Remark on the DGLAP-fits for structure functions}

The asymptotics (\ref{as}) is of the Regge type. The Regge behavior is generated by total tesummations of
leading logarithms of $x$ and it is unrelated to behavior of
the initial parton densities $\delta q (x)$ and $\delta g (x)$. In particular,
both $\delta q (x)$ and $\delta g (x)$ are not supposed to reveal a
power  behavior $\sim x^{-a}$ at small $x$. On the contrary, the standard DGLAP fits
\begin{equation}\label{fit}
\delta q, \delta g = N x^{-a} (1 - x)^b (1 + cx)^d,
\end{equation}
with $a,b,c,d,N > 0$,
conventionally includes such factors. The singular factors $x^{-a}$ are incorporated in the fits ad hoc in order to provide
$f$ with a steep growth at small $x$ and thereby meet experimental data. Indeed, in the $\omega$ -space
the factor $x^{-a}$ corresponds to the pole contribution
\begin{equation}\label{pole}
\delta q_p = 1/(\omega - a).
\end{equation}
Being installed in Eq.~(\ref{fgen}), the pole contribution (\ref{pole}) straightforwardly leads to
the Regge asymptotics

\begin{equation}\label{asa}
f \sim x^{-a} \left[\frac{\ln \left(Q^2/\Lambda^2\right)}{\ln \left(\mu^2/\Lambda^2\right)}\right]^{\gamma (a)/b},
\end{equation}
which differs a lot from both Eq.~(\ref{as}) and the well-known LO DGLAP -asymptotics

\begin{equation}\label{asdglap}
  f \sim \exp \sqrt{\frac{C}{2 \pi b}\ln(1/x)\ln \left[\frac{\ln \left(Q^2/\Lambda^2\right)}{\ln \left(\mu^2/\Lambda^2\right)}\right]}.
\end{equation}
Let us note that Eqs.~(\ref{as},\ref{asdglap}) involve non-singular fits.
Confronting (\ref{as}) to (\ref{asa}) proves that the only role of the singular factors $x^{-a}$ is,
actually, to mimic the total resummation of $\ln^n x$.  When the resummation is accounted for, these
factors can be dropped and therefore the number of parameters in the fits can be reduced.
The qualitative difference between the asymptotics
(\ref{as}) and (\ref{asa}) includes two items:

\textbf{(i)} The intercept $\Delta$ in Eq.~(\ref{as}) is
calculated while $a$ in (\ref{asa}) is fixed from experiment.

\textbf{(ii)} The factor $x^{- \Delta}$ in Eq.~(\ref{as}) is present in the asymptotic expressions, i.e. at $x \to 0$,
only. It never appears in expressions for the structure functions like
Eq.~(\ref{fgen}) at finite $x$. On the contrary as soon as the fit (\ref{fit}) is used, the factor
$x^{-a}$ is present in Eq.~(\ref{fgen}) at any $x$, including large $x \sim 1$. It means that, in a sense, the $x$ -dependence
of $f$ is always controlled by its asymptotics, when singular fits are used. This contradicts to the observation made in
Ref.~\cite{egtdglap}: the small-$x$ asymptotics represent the structure functions reliable at very
small $x$ ($x < 10^{-8}$) only while at $x > 10^{-8}$ the structure function is much greater than
its asymptotics. The latter leads to the quantitative difference between $a$ and $\Delta$:
In order to force Eq.~(\ref{asa}) to represent $f$ at presently available $x$, i.e. at $x > 10^{-8}$, one has to increase $a$, which leads to
the following relation:

\begin{equation}\label{adelta}
  a > \Delta .
\end{equation}

By this reason, we will name the fictitious intercepts $a$ as pseudo-intercepts. As their values
are determined from experiment at available $x$, they always exceed genuine intercepts.
So, the situation looks like that: On one hand, the use
of the pseudo- intercepts $a$ allows one to approximate Eq.~(\ref{fgen}) by its asymptotics which is given by simple
Regge-like expressions (\ref{asa}). On the other hand, the power $x$ -dependence in Eq.~(\ref{asa})  has nothing to do with
impact of genuine Reggeons introduced in Theory of the Regge poles.
The wide-spread tactics to use pseudo-Reggeons, instead of resummation of $\ln x$,
in order to solve immediate practical tasks
leads to serious theoretical problems especially important for the
singlet structure function $F_1$ and the singlet parton distributions: there are no available
expressions for those objects besides their Regge asymptotics $x^{-\Delta_P}$, with $\Delta_P$ being the Pomeron
intercept. In order to fit such asymptotics to explanation of experimental data at available energies,
they use one or several Pomerons with large pseudo-intercepts $\Delta_P$ violating the Froissart bound.
Finally, let us note that necessity to include singular factors
in the fits is a clear indication that essential logarithms of $x$ are not accounted for.

\section{Structure functions at small $Q^2$}

Description of the structure functions at small $Q^2$ is important because this kinematics has
been investigated experimentally. For instance, the spin-dependent structure functions at small $Q^2$
 are investigated by the COMPASS Collaboration (see e.g. Ref.~\cite{compass}). On the other hand, this region is
 absolutely beyond the reach of DGLAP. Despite this, sometimes in the literature
(see e.g. Refs.~\cite{bad,zotov}, recent paper Ref.~\cite{kotik} and refs therein)
the DGLAP -parametrization $\alpha_s = \alpha_s (Q^2)$  is treated as an exact
expression where $Q^2$ can acquire an arbitrarily small values and therefore the shift

\begin{equation}\label{ashift}
\alpha_s (Q^2) \to \alpha_s (Q^2 + \mu^2)
\end{equation}
is needed at small $Q^2$ and especially at $Q^2 \to 0$ in order to keep $\alpha_s$ within the
 perturbative domain.
However,  in the present paper we have shown that the DGLAP-parametrization
$\alpha_s = \alpha_s (Q^2)$ is an approximation valid for large $Q^2$ only, where the shift (\ref{ashift})
is totally unnecessary and even cannot be seen at $\mu^2 \ll Q^2$.
Actually, the DGLAP -parametrization fails at small $Q^2$ or at small $x$ , so $\alpha_s (Q^2)$ should be replaced by the effective
coupling $\alpha_{eff}$. The coupling $\alpha_{eff}$ does not depend on $Q^2$
 at small $Q^2$ or small $x$ at all, which
makes the shift (\ref{ashift}) absolutely unnecessary.
Let us stress that interpretation of the approximation $\alpha_s = \alpha_s (Q^2)$ and the shift (\ref{ashift})
as exact expressions has led to various misconceptions abundant in the literature. On the other hand,
the small-$Q^2$ kinematics has been investigated experimentally by the COMPASS Collaboration (see e.g. Ref.~\cite{compass}),
so it is important to describe the structure functions in this region. In
Ref.~\cite{egtsmq} we proved that the expressions in Ref.~\cite{egtsns} for the structure function $g_1$ at small $x$ and
large $Q^2$ can be extended to small $Q^2$ by shifting
\begin{equation}\label{kshift}
 k_{\perp}^2 \to k_{\perp}^2 + \mu^2
\end{equation}
in propagators of the soft quarks and gluons.
The shift (\ref{kshift}), where $\mu$ is a
cut-off,  is necessary to regulate infrared singularities in involved Feynman graphs.
This shift eventually leads to the shift
\begin{equation}\label{shift}
Q^2 \to Q^2 + \mu^2
\end{equation}
in logarithmic contributions to the structure functions.
Our estimate for $\mu$ obtained with using the
Principle of Minimal Sensitivity~\cite{pms} was $\mu \approx 10 \Lambda_{QCD}$ (for more
details see our recent paper \cite{egtfrozen}).  The
shift (\ref{shift}) can be neglected at $Q^2 >> \mu^2$ but becomes essential at small $Q^2$,
making possible description of the kinematic region where $Q^2$ are small. For example, DIS at such a
kinematics has been studied experimentally by the COMPASS collaboration.
Therefore, it is the shift (\ref{kshift}) that brings theoretical grounds for description of the
small $Q^2$ -region. Let us stress that this shift should not been applied to $\alpha_s$ and
it has nothing to do with the baseless modification (\ref{ashift}) of the coupling. On the other hand,
the shift (\ref{kshift}) together with the shift
$\alpha_s (k^2_{\perp}) \to \alpha_s (k^2_{\perp} + \mu^2)$ can be used
when $k$ are momenta of the virtual $t$-channel gluons connecting the perturbative part of the
structure function to the parton distributions but these shifts have nothing to do with the shift of
Eq.~(\ref{ashift}) (see Ref.~\cite{egtfrozen}
for details).

\section{Remark on higher twists}

It is interesting to notice that the shift (\ref{ashift}) can be used to clarify the problem of
higher twists and, in addition, it explains the puzzle
of the power $Q^2$ -corrections. A namely, the contributions $\sim 1/(Q^2)^n$ to DIS
structure functions, usually attributed to higher twists, are known to be present
at large $Q^2$ but enigmatically disappear at small $Q^2$ where they could be
extremely impactive. A simple and natural solution of this puzzle was found in  Ref.~\cite{egtsmq} (see also
overview~\cite{g1sum}). In brief, it can be reduced to the following: Among different contributions
to the structure functions, there are contributions

\begin{equation}\label{tq}
 T_n \sim \ln^n \left((Q^2 + \mu^2)/\mu^2\right).
\end{equation}

In the region of large $Q^2$, where $Q^2 > \mu^2$, such contributions can be expanded into power series in the
following way:

\begin{equation}\label{tbigq}
 T_n \sim \ln^n \left(Q^2/\mu^2\right) + \sum \left(\frac{\mu^2}{Q^2}\right)^k.
\end{equation}

The logarithmic term in Eq.~(\ref{tbigq}) is included into the leading twist contributions
whereas the power terms are identical to the power terms  attributed to
higher twists. Such terms are conventionally supposed to have non-perturbative
origin. However, Eq.~(\ref{tbigq}) explicitly demonstrates that there are
the power terms of purely perturbative nature. Surely, they should be accounted for
in the first time and only the rest should be attributed to higher twists.

On the other hand, the expansion (\ref{tbigq}) holds at large $Q^2$ only.
In the small-$Q^2$ region the power expansion of Eq.~(\ref{tq}) looks different:

\begin{equation}\label{tsmallq}
 T_n \sim  1 + \sum \left(\frac{Q^2}{\mu^2}\right)^k.
\end{equation}

The first term in the r.h.s. of Eq.~(\ref{tsmallq}) is included into the leading twist
contributions while the other terms are again the power-$Q^2$ corrections, though
with $Q^2$ in the nominator. Eq.~(\ref{tsmallq})
proves that the power corrections $\sim 1/(Q^2)^n$ can never become singular at
small $Q^2$. Comparing Eq.~(\ref{tbigq}) to (\ref{tsmallq}) reveals that the
corrections $\sim 1/(Q^2)^n$ are about vanishing at

\begin{equation}\label{q0}
Q^2 \sim \mu^2,
\end{equation}
where none of expansions (\ref{tbigq},\ref{tsmallq}) can be used. In Ref.~\cite{egtsns}
we fixed $\mu$ for the non-singlet structure
functions $F_1^{NS}$ and $g_1^{NS}$:
for them $\mu_{NS} \sim 1$~GeV. Our prediction of vanishing power corrections  $\sim 1/(Q^2)^n$
at values of $Q^2$ approaching $\sim 1$~Gev$^2$ perfectly agrees with results
obtained by analysis of experimental data.

\section{Conclusion}

In the present paper we have analyzed the $Q^2$ -dependence of the structure functions, considering
separately this dependence in the kinematic region of large and small $x$.  Our consideration
embraced both large and small values of $Q^2$. We
argued against the conventional point of view that the $Q^2$-dependence
at any $Q^2$ and $x$ is originated by the DGLAP -parametrization $\alpha_s = \alpha_s (Q^2)$ leading to
$Q^2$- dependence of the intercepts of the structure functions. We showed that this
parametrization holds
exceptionally in the region of large $x$ and $Q^2$, providing that in this region the factorization scale
should be large. Outside this region, $\alpha_s$ should be replaced by the effective coupling $\alpha_{eff}$.
At small $x$, coupling $\alpha_{eff}$ depends on both transverse and longitudinal momenta,
 which destroys the factorization of the phase space into the longitudinal and transverse spaces taking
 place in DGLAP. In addition, the integration limits do not involve $Q^2$.
 As a result, the effective coupling does not depend on $Q^2$ in the small-$x$ region.
 This leads to the Regge small-$x$ asymptotics
of the structure functions, with intercepts independent of $Q^2$. We also argued against the shift
$\alpha_s (Q^2) \to \alpha_s (Q^2 + \mu^2)$  used in the literature to keep $\alpha_s$
in perturbative domain at small $Q^2$. In this regard, we reminded that the parametrization $\alpha_s (Q^2)$
as well as DGLAP in whole should be used at large $Q^2$ only.
Superficial use of the shift  $\alpha_s (Q^2) \to \alpha_s (Q^2 + \mu^2)$ at small $Q^2$ has led
to various misconceptions known in the literature. In contrast, we
advocated the use of shift $Q^2 \to Q^2 + \mu^2$
in such evolution equations, derivation of which is
not based on assuming $Q^2$ large. We stressed that the shift did not involve parametrization
of the QCD coupling. Eventually, we have
considered theoretical grounds for such a shift and discussed its application to the power $Q^2$-
contributions to the structure functions.

\section{Acknowledgement}

The work is partly supported by Grant RAS 9C237,
Russian State Grant for Scientific School RSGSS-65751.2010.2 and
EU Marie-Curie Research Training Network under contract
MRTN-CT-2006-035505 (HEPTOOLS).

\end{document}